\documentstyle[aps,12pt,epsf]{revtex} \bibstyle{unsrt}
  \large \normalsize

\begin{document}

\begin{center}
\vspace*{1.0cm}

{\Large\bf Coherent optimal control of multiphoton molecular excitation}

\vspace{1.0cm}

Bijoy K Dey

\vspace*{0.2cm}
{\small \sl Department of Chemistry, University of Toronto, Toronto, Ontario, Canada M5S 3H6\\email:bdey@tikva.chem.utoronto.ca}
\end{center}

\newpage

\begin{center}

\vspace*{0.7cm}

{\bf Abstract}
\end{center}
~~~~~~~~~~We give a framework for molecular multiphoton excitation process induced by an optimally designed electric field. The molecule is initially prepared in a coherent superposition state of two of its eigenfunctions. The relative phase of the two superposed eigenfunctions has been shown to control the optimally designed electric field which triggers the multiphoton excitation in the molecule. This brings forth flexibility in designing the optimal field in the laboratory by suitably tuning the molecular phase and hence by choosing the most favorable interfering routes that the system follows to reach the target. We follow the quantum fluid dynamical formulation for designing the electric field with application to HBr molecule.

\vspace*{1.0cm}
\section{Introduction}
~~~~~~~~~~Finding ways to control the quantum phenomena is, arguably, the essence of chemical and physical processes. Advancement in generating tailored light pulses has made it possible to steer matter towards a specific goal, namely, to control its future \cite{ref1,ref2,ref3,ref4,ref5,ref6,ref7}. It is now well established that light shaping is an efficient way of controlling quantum systems and thus, for example, chemical reactions. Recent days have suggested a variety of control schemes and their implementations in the laboratory \cite{ref8,ref9,ref10,ref11,ref12,ref13,ref14} based on the coherent nature of the laser radiation.

~~~~~~~~~~Coherent control (CC) scheme accesses the broad range of quantum interference effect through the relative phase of the light waves and/or the relative phase of the molecular eigenfunctions as has been theoretically demonstrated by Brumer and Shapiro \cite{ref5,ref6} and experimentally realized by others \cite{ref14}. A somewhat different control scheme uses the rapid progress in ultrashort laser pulse technology, e.g. , Tannor et al \cite{ref2,ref15} suggested a ``pump-dump'' technique, which has been realized experimentally by several groups \cite{ref16,ref17}. These techniques deliver control through a limited number of parameters, viz., the optical phase, the molecular phase and the time delay between pump and dump laser pulses.

~~~~~~~~~~In the general field of controlling dynamics these parameters are not sufficient and it was Rabitz et. al \cite{ref18} who first proposed in their optimal control theory (OCT) that the electric field of the laser pulse should be treated as parameter and be specifically designed both temporally and spectrally. Theoretically, OCT creates a nonstationary state of one's choice, by optimally designing the control field. The control field couples the system with itself in a way that as the system evolves, its motion is modified by the influence of the field along its path, and the optimal control field provides the maximum overlap of the time-evolved state with the state of one's choice. Although there have been progress in the pulse-shaping technology, it still remains a huge task ahead to design such field in the laboratory.

~~~~~~~~~~In this paper, we present a theoretical treatment of the OCT which introduces the parameter, molecular phase, mostly used in the CC scheme. This prepares the molecule in a coherent superposition state comprising of two molecular eigenfunctions, prior to its being submitted to the control field. The molecular phase is shown to be an experimental parameter which can take a whole range of values for designing the control field. We organize this paper as follows : section 2 describes the preparation of the initial superposition state comprising of two vibrational eigenfunctions of the ground electronic state of HBr molecule. This can be accomplished by non-resonant two-photon interaction between the molecule in the ground state and two laser electric fields. The relative phase of the laser fields defines the molecular phase of the eigenfunctions forming the superposition. Section 3 briefly presents OCT within the framework of quantum fluid dynamics (QFD), called OCT-QFD method, for desiging an electric field to meet a specific objective. The details of the OCT-QFD have been published before \cite{ref19,ref20}. Section 4 applies the OCT-QFD for the manipulation of the multiphoton excitation of HBr molecule subject to the design of an optimal control field. Section 5 concludes the paper.

\section{Molecular coherence and Superposition state}
Here we showed one possible way to prepare the superposition state of HBr. The approach is based on a nonresonant two-photon transition phenomena whereby the molecule intially in the eigenstate $|\nu_1J_1M_1>$ transfers population to the eigenstate $|\nu_2J_2M_2>$ through the intermediate states. Here $|\nu JM>$ refers to the ro-vibrational eigenfunction corresponding to an electronic state, where $\nu$ is the vibrational quantum number, J the rotational quantum number and M the projection of J onto the internuclear axis in the laboratory coordinate. Let us consider the superposition where both $|\nu_1J_1M_1>$ and $|\nu_2J_2M_2>$ belong to the same electronic state, viz., the ground state $^1\Sigma^+$ of HBr molecule. The two-photon excitation is assumed take place by the laser field given by

\begin{eqnarray}
E(t)=\hat{\eta} E^{(0)}f(t)cos\omega_L t(1+cos\phi_L)
\end{eqnarray}
where
\begin{eqnarray}
f(t)=e^{-(\frac{t-t_0}{\tau})^2}
\end{eqnarray}
with $t_0$ being the switch-on time of the pulse, $\tau$ being related to the full width at half maximum(FWFM) as $FWHM=2\tau\sqrt{-ln(1/2)}$, $\omega _L$ the central frequency of the pulse and $E_i^{(0)}$ the maximum amplitude of the laser electric field. The electric field polarization vector $\hat{\eta}$ is considered along z direction in the laboratory coordinate. This field can be obtained by the superposition of two identical laser pulses with relative phase $\phi_L$. The superposition state is thus given by

\begin{eqnarray}
\Psi_s(t)=c_1(t)|\nu_1J_1M_1>e^{-i\omega_1t}+c_2(t)|\nu_2J_2M_2>e^{-i\omega_2t}
\end{eqnarray}
At $t=-\infty$ $|c_1|^2=1$ and at time t $|c_1(t)|^2=1-|c_2(t)|^2$. In the above equation $\omega_i=\frac{E_i}{\hbar}$, where $E_i$ is the energy of the $|\nu_iJ_iM_i>$ eigenstate. For two-photon interaction, applying the second order time-dependent perturbation theory we obtain

\begin{eqnarray}
c_2(t)=-\frac{1}{\hbar^2}c_1\sum_Id_{I1}^{\eta}d_{2I}^{\eta}\int_{-\infty}^te^{i(\omega_{2I}-i(\Gamma_2+\Gamma_I)/2)t^\prime}E(t^\prime)G(t^\prime)dt^\prime
\end{eqnarray}
where $G(t^\prime)$ is given as
\begin{eqnarray}
G(t^\prime)=\int_{-\infty}^{t^\prime}e^{i(\omega_{I1}-i(\Gamma_I+\Gamma_1)/2)t^{\prime\prime}}E(t^{\prime\prime})dt^{\prime\prime}
\end{eqnarray}

I refers to the intermediate state involved in the two-photon absorption process, $\Gamma_i$'s are the life time of the ith state and $d_{ij}^\eta $ is the dipole matrix elements between i and j states. The summation involving only the dipole matrix elements in Eq.(4) can be written as \cite{ref21}
\begin{eqnarray}
\sum_Id_{I1}^{\eta}d_{2I}^{\eta}=\sum_{\nu_{I}} (-1)^{M+M_1-K-K_1}F_{\nu_2\nu_I}\\\nonumber
&&F_{\nu_I\nu_1}(2J_1+1)^{1/2}(2J_2+1)^{1/2}L_M^{(J)}(\eta)\\\nonumber
&&M_K^{(J)}(\mu^e_{gI})\left(\begin{array}{ccc}J_1 & J & J_2 \\-M_1 & -M & M_2 \end{array}\right)\left(\begin{array}{ccc}J_1 & J & J_2 \\-K_1 & -K & K_2 \end{array}\right)
\end{eqnarray}
The function $F_{\nu_{I}\nu_{1}}$ is the Frank Condon factor whereas the term $L_M^{(J)}(\eta)$ is purely geometric depending on the direction of the polarization vector of the laser field and is given by

\begin{eqnarray}
L_M^{(J)}(\eta)&=&(2J+1)^{1/2}\sum_{A,B}(-1)^{A+B}\eta_A\eta_B\\\nonumber
&&\left(\begin{array}{ccc}1 & 1 & J \\-A & -B & M \end{array}\right)
\end{eqnarray}
where A and B run over -1, 0 and 1 corresponding to X, Z and Y respectively in the laboratory coordinate. The term $M_K^{(J)}(\mu^e_{gI})$ is free from the experimental conditions which is purely molecular and is given by
\begin{eqnarray}
M_K^{(J)}(\mu^e_{gI})&=&(2J+1)^{1/2}\sum_{a,b}(-1)^{a+b}\\\nonumber
&&\left(\begin{array}{ccc}1 & 1 & J \\-a & -b & K \end{array}\right)(\mu^e_{gI})_a(\mu^e_{Ig})_b
\end{eqnarray}
where a and b run over -1, 0 and 1 corresponding to x, z and y respectively in the molecular coordinate. The function $(\mu^e_{gI})_a=(\mu^e_{Ig})_a$ is the a-th component of the elctronic dipole matrix element between the ground electronic state and the intermediate states. It is obvious that the intermediate state must be either $\Sigma^+$ (when the transition dipole operator is $-\sum_lz_l$ \cite{ref22}) or $\Pi$ (when the dipole operator is $\frac{1}{\sqrt{2}}\sum_l(x_l\pm iy_l)$ \cite{ref22}). Whereas $J_1$, $J_2$, J, $K_1$, $K_2$, K, $M_1$, $M_2$ and M take the values as K=0, M=0, $M_2=M_1$, $K_1=K_2=0$, $J_2=J_1$ (or $J_2=J1+2$). The vibrational quantum numbers for the superposition can be either $\nu_2=\nu_1$ or $\nu_2\ne\nu_1$.  Note that the assumption of the perturbation theory implies that $|c_2(t)|^2$ cannot exceed $\equiv 0.2$ which, in other words, restricts the power of the pump laser \cite{ref23}. Notice that the molecular phase defined as $\phi_M=tan^{-1}(\frac{\Im(c_2)}{\Re(c_2)}$, where $\Re$ and $\Im$ refer to the real and the imaginary parts respectively, depends mainly on $\phi_L$, $\omega_1$ and $\omega_2$. The factor $(\Gamma_I+\Gamma_2)/2$ or $(\Gamma_I+\Gamma_1)/2$ in the exponent of equations 4 and 5 respectively, have the least contribution to the molecular phase since they are neglible compared to $\omega_{2I}=\omega_2-\omega_I$ or $\omega_{I1}=\omega_I-\omega_1$ respectively. Thus one can vary $\phi_M$ in the laboratory by varying $\phi_L$, $\omega_1$ and $\omega_2$.

\section{OCT-QFD method}

In optimal control theory an objective functional, corresponding to a specific dynamics of one's choice, is minimized with respect to the electric field by solving the time-dependent Schroedinger (TDSE) equation. Consider a general target expectation value defined as 
\begin{eqnarray}
\Theta_T=\int_o^T\Theta\rho(x,T)dx
\end{eqnarray}
where $\Theta$ is an observable operator and $\rho(x,T)=\Psi^*(x,T)\Psi(x,T)$ with $\Psi(x,t)$ being the complex wave function at the target time t=T. This wave function $\Psi(x,t)$ obeys the TDSE
\begin{eqnarray}
i\hbar\frac{\partial\Psi(x,t)}{\partial t}=[-\frac{\hbar^2}{2m}\nabla ^2 +V(x) + V_{ext}(x,t)]\Psi(x,t)
\end{eqnarray}
where V typically confines the particle in a locale and $V_{ext}$ is the control taken here as $-\mu (x)E_c(t)$ with $E_c(t)$ being the control electric field to be designed and $\mu (x)$ the dipole moment.

The goal is to steer $\Theta_T$ as close as possible to a desired value $\Theta^d$. The active spatial control interval is taken as $x_l <x <x_r$ over the time $0<t<T$ that the control process occurs. We now desire to minimize the cost functional $J_{cost}=J_{target}+J_{field}$, where $J_{target}$ and $J_{field}$ are given by
\begin{eqnarray}
J_{target}=\frac{1}{2}\omega_x(\Theta_T-\Theta^d)^2
\end{eqnarray}
and
\begin{eqnarray}
J_{field}=\frac{1}{2}\omega_e\int_o^TE_c^2(t)dt
\end{eqnarray}
The minimization of $J_{cost}$ with respect to $E_c(t)$ must be subjected to the satisfaction of the equations of motion for $\Psi(x,t)$ in Eq.(10), which can be transformed into two equations, viz., the continuity equation
\begin{eqnarray}
\frac{\partial \rho}{\partial t}+\nabla \cdot (\rho {\bf v})=0
\end{eqnarray}
and a modified Hamilton Jacobi equation
\begin{eqnarray}
\frac{\partial S}{\partial t}+\frac{\nabla S\cdot \nabla S}{2m}+V+V_{ext}+V_q=0
\end{eqnarray}
with the substituition $\Psi(x,t)=\rho^{1/2}(x,t)e^{iS(x,t)/h}$ in Eq.(10) where $V_q=-\frac{\hbar ^2}{2m}\frac{\nabla ^2\rho ^{1/2}}{\rho ^{1/2}} = -\frac{\hbar ^2}{2m}[\nabla ^2ln\rho ^{1/2}+(\nabla ln\rho ^{1/2})^2]$. This forms the basis of the QFD\cite{ref24,ref25} treatment of TDSE. This equation can be transformed into the one for the evolution of the velocity vector {\bf v} by taking the gradient to give 
\begin{eqnarray}
\frac{\partial}{\partial t}{\bf v}=-({\bf v}\cdot \nabla ){\bf v}-\frac{1}{m}\nabla (V+V_{ext}+V_q)
\end{eqnarray}
Defining the quantum current as
\begin{eqnarray*}
{\bf j}({\bf x},t)=-\frac{\hbar}{m}\Im[\Psi ^*({\bf x},t)\nabla \Psi ({\bf x},t)]=\rho ({\bf x},t){\bf v}({\bf x},t),
\end{eqnarray*}
one readily obtains the equation of motion for {\bf j} as
\begin{eqnarray}
\frac{\partial }{\partial t}{\bf j}=-{\bf v}(\nabla \cdot {\bf j})-({\bf j}\cdot \nabla ){\bf v}-\frac{\rho}{m}(\nabla V+V_{ext}+V_q)
\end{eqnarray}

Thus within the QFD formulation, we need to minimize $J_{cost}$ with respect to $E_c(t)$ subject to the satisfaction of the equations of motion for $\rho$ and {\bf j} given by Eqs.(13) and (16) respectively.

We may fulfill this constraint by introducing the unconstrained cost functional as 
\newpage
\begin{eqnarray}
\bar{J}&=&J_{cost}-\int_0^T\int_{x_{l}}^{x_{r}}\lambda _1(x,t)[\frac{\partial \rho(x,t)}{\partial t}+\frac{\partial j(x,t)}{\partial x}]dx dt\\\nonumber
&&-\int_0^T\int_{x_{l}}^{x_{r}}\lambda _2(x,t)[\frac{\partial j(x,t)}{\partial t}+\frac{\partial }{\partial x}(\frac{j^2}{\rho })+\frac{\rho}{m}\frac{\partial }{\partial x}(V+V_q+V_{ext})]dx dt
\end{eqnarray}  
where $\lambda _1(x,t)$ and $\lambda _2(x,t)$ are Lagrange's multiplier functions. \\

An optimal solution satisfies $\delta \bar{J}=0$, which is assured by setting each of the functional derivatives with respect to $\lambda _1$, $\lambda _2$, $\rho $, {\bf j} and $E_c$ to zero. The first two, i.e., the functional derivatives with respect to $\lambda _1$ and $\lambda _2$ regenerate the QFD equations viz., Eqs.(13) and (16). The three others are obtained in the forms :

\begin{eqnarray}
\frac{\partial \lambda _2}{\partial t}+\frac{\partial}{\partial x}(\lambda _2v_{\lambda})+S_1[\rho ,j,\lambda _2]=0
\end{eqnarray}

\begin{eqnarray}
\frac{\partial \lambda _1}{\partial t}+\frac{\partial}{\partial x}(\lambda _1v_{\lambda})-\lambda _2\frac{\partial}{\partial x}(V+V_q(\lambda _2)+V_{ext})+S_2[\rho ,j,\lambda _2]=0
\end{eqnarray}
and
\begin{eqnarray}
\frac{\delta \bar{J}}{\delta E_c(t)}=\int_{x_{l}}^{x_{r}}\lambda _2(x,t)\rho (x,t)\frac{\partial }{\partial x}\mu (x)dx +\omega _eE_c(t)=0
\end{eqnarray}
where
\begin{eqnarray}
S_1=2\frac{j}{\rho }\frac{\partial \lambda _2}{\partial x}
\end{eqnarray}

\begin{eqnarray}
S_2&=&-\frac{\lambda _2}{m}\frac{\partial}{\partial x}(V_q(\rho)-V_q(\lambda _2))-\frac{j^2}{\rho ^2}\frac{\partial \lambda _2}{\partial x}\\\nonumber
&&-\frac{\hbar ^2}{4m^2\rho ^{1/2}}\frac{\partial ^2}{\partial x^2}[\frac{1}{\rho ^{1/2}}\frac{\partial}{\partial x}(\lambda _2\rho )]\\\nonumber
&&+\frac{\hbar ^2}{4m^2\rho ^{3/2}}\frac{\partial ^2}{\partial x^2}\rho ^{1/2}\frac{\partial}{\partial x}(\lambda _2\rho )
\end{eqnarray}
and
\begin{eqnarray}
V_q(\lambda _2)=-\frac{\hbar ^2}{2m}\frac{\nabla ^2\lambda _2 ^{1/2}}{\lambda _2 ^{1/2}} = -\frac{\hbar ^2}{2m}[\nabla ^2ln\lambda _2 ^{1/2}+(\nabla ln\lambda _2 ^{1/2})^2]
\end{eqnarray}

The corresponding final conditions are
\begin{eqnarray}
\omega _x[\Theta_T-\Theta ^d]\Theta (x)-\lambda _1(x,T)=0
\end{eqnarray}
and
\begin{eqnarray}
\lambda _2(x,T)=0
\end{eqnarray}

The equations (18) and (19) for $\lambda _2$ and $\lambda _1$ respectively ressemble that of $\rho $ and {\bf j} with the only difference being the extra source terms $S_1$ and $S_2$. The source terms depend on $\rho $ and {\bf j}. $v_{\lambda}$ in the above equations is the 'velocity' associated with the Lagrange's multiplier and is given as $v_{\lambda}=\frac{\lambda _1}{\lambda _2}$. There are now two different quantum potential terms, one of which is a function of $\rho (x,t)$ and the other is a function of $\lambda _2(x,t)$. In this formalism the evolution of $\lambda _1(x,t)$ takes place by $V_q(\lambda _2)$ as well as the difference of the two types of quantum potential. In obtaining the above equations we have standardly assumed no variation of either $\rho (x,0)$ or j(x,0). Thus, we start from the initial value of $\rho (x,0)$ and j(x,0) to solve Eqs.(13) and (16). Eqs.(18) and (19) can be solved for $\lambda _2(x,t)$ and $\lambda _1(x,t)$ by integrating backward from time T using  $\lambda _1(x,T)$ and $\lambda _2(x,T)$ given in Eqs.(24) and (25) respectively. The equations (13), (16), (18) and (19) are non-linear thereby calling for iteration to solve(see ref. \cite{ref24} and \cite{ref25} for details). Finally the desired control electric field is given from Eq.(20) as 
\begin{eqnarray}
E_c(t)=-\frac{1}{\omega _e}\int_{x_{l}}^{x_{r}}\lambda _2(x,t)\rho (x,t)\frac{\partial }{\partial x}\mu (x)dx
\end{eqnarray}

\section*{4 Application to HBr Molecule}

The OCT-QFD has been applied for manipulating the multiphoton excitation process of HBr molecule whose initial density $\rho(x,0)$ is given by 
\begin{eqnarray}
\rho(x,0)=|c_1|^2\rho _1(x,0)+|c_2|^2\rho _2(x,0)+4|c_1||c_2|cos\phi_M\rho_{12}(x,0)
\end{eqnarray}
where $c_1$ and $c_2$ have been obtained following section 2. In Eq.(27) $\rho_i(x,0)=|\psi_i(x,0)|^2$ and the $\rho_{12}(x,0)=\psi_1(x,0)\psi_2(x,0)$ where $\psi_i(x,0)$ corresponds to the $|\nu_iJ_jM_i>$ eigenstate with x being the internuclear distant. Although the perturbation theory permits $|c_2|^2\le0.2$ there could be other non-perturbative methods resulting $|c_2|^2>0.2$. Thus, in the results below whenever we consider $|c_2|^2>0.2$ we assume the existence of some non-perturbative methods. A whole range of $\phi_M$ can be attempted to follow the different interfering routes to reach to the target state by suitably modifying the control field $E_c(t)$.

Equation (4) for $c_2$ shows that we need the ro-vibrational eigenfunctions in the ground and the excited electronic states. These have been evaluated by solving the time independent Schr\"{o}dinger equation using the Fourier Grid Hamiltonian(FGH) method \cite{ref26}. The ground and the excited electronic potentail energies have been taken from the ref.\cite{ref27}. Figure 1 shows $c_2$ for the superposition $c_1|000>e^{-iE_1t/\hbar}+c_2|100>e^{-iE_2t/\hbar}$ where $E_1$ and $E_2$ are the energies of the eigenstates $|000>$ and $|100>$ respectively. This gives $|c_2|^2=0.2$, $|c_1|^2=0.8$ and $\phi _M= 0.25$ radian after the laser fields disappear. For any superposition one can choose a whole range of values for $\phi _M$ for a given $|c_1|^2$ and $|c_2|^2$ by choosing different values of $\phi _L$. This $\phi _M$, $\omega_1$, $\omega_2$, $|c_1|$ and $|c_2|$ have been used as parameters for designing the control field for the occurrence of a specific dynamical change in the molecule corresponding to the target operator $\Theta = x$, where x represents the average distant along the internclear axis. Other kinds of target operator can also be considered in the present QFD-OCT following some modifications of the control equations (see ref. \cite{ref20}). In the results shown below the spatial range of the calculation was $0 \le x(a.u.) \le 14$ and the time interval was $0 \le t \le T$ with T=5000 a.u. The total number of the spatial mesh points taken was 52 which gives $\delta x = 0.27$ a.u. whereas the total number of time steps was 2000 which gives $\delta t= 2.5$ a.u. The weight $\omega_e$ is taken 0.5 and $\omega_x$ 500 and the desired target value $\Theta ^d=3.2$ a.u.

Figures 2,3,4 and 5 show the results for the optimal control field and the dynamics involved. Notice that the desired target, i.e., $\Theta ^d=3.2 $ a.u. is the same for alll the cases (Figs. 3, 4(b) and 5((b),(d))) although the paths through which the dynamics occur to reach the desired target were different. Basically, the phenomena that causes the expectation value of x to change in time is the multi-photon excitation of the molecule induced by the control field along a selective path. If we assume that the control field strengths are low enough for the perturbation theory to be valid, which is indeed true in the present results(e.g., $E_{max}\approx 0.3$ a.u.(Fig.2(c)) corresponding to maximum intensity $\approx 10^{14} Watt/cm^2$ and $E_{min}\approx 0.06$ a.u.(Fig.5(c)) corresponding to $\approx 10^{13} Watt/cm^2$ intensity) we find that the time evolved wave function under the control field is
\begin{eqnarray}
\Psi(t)=\Psi_s(t)+\sum_j c_j(t)\psi_je^{-i\omega_jt}
\end{eqnarray}
where the summation j occurs over all the possible states excited by the optimal field starting from the superposition state. In the above equation $c_j(t)$ depends on $c_1$ and $c_2$ as $c_j(t)=c_1 f_1(t)+c_2 f_2(t)$, e.g., in the first order perturbation theory $f_i$, is given as

\begin{eqnarray}
f_i&=&-\frac{1}{\hbar}d_{ji}\int_0^te^{i(\omega_{ji}-i(\Gamma_j+\Gamma_i)/2)t^\prime}E(t^\prime)dt^\prime
\end{eqnarray}

where E(t) is the optimal control field. Thus, the density at any time after the control field is on is given by $\rho(x,t)=\rho _{ni}(x,t)+\rho _{in}(x,t)$ where ``in'' referes to the interference term and ``ni'' to the non-interference term. Similarly the target expectation has both non-interference and interference terms in it, i.e., $<x>(t)=<x>_{ni}(t)+<x>_{in}(t)$. This readily gives the optimal control field $E(t)=E_{ni}(t)+E_{in}(t)$ The non-interference terms associated with $\rho_{ni} (x,t)$, $<x>_{ni}(t)$ or $E_{ni}(t)$ have two parts, one relates to $|c_1|^2$ and the other to $|c_2|^2$ whereas the interference terms depend on $|c_1|$, $|c_2|$ and $\phi _M$.

Thus, one actually controls the dynamics of the multiphoton excitation process subject to the suitable designing of an optimal electric field which itself can be controlled by varying the parameters, viz., $\phi _M$, $|c_1|$, $|c_2|$, $\omega_1$ and $\omega _2$. This has been depicted through the figures 2 to 5. The dynamics by the control field, take place through three routes : (a) route that ends up at $\psi _j$ (Eq.(28)) from $|\nu_1J_1M_1>$, the probability of which is proportional to $|c_1|^2$, (b) route that ends up at $\psi _j$ from $|\nu_2J_2M_2>$, the probability of which is proportional to $|c_2|^2$ and (c) the route that end up at $\psi _j$ through the interference between the routes (a) and (b), the probability of which is proportional to $|c_1||c_2|sin\phi _M$ or $|c_1||c_2|cos\phi _M$. Thus the molecular phase can cause certain excitations by the control field during the process of its designing, which are not present in the absence of the superposition state.
 
Figure 2 shows the control fields corresponding to four different values of $\phi _M$ for the superposition $|000>+|100>$. These fields excite several vibrational excited states(not shown here) in the process of achieving the target state, i.e., a state with $\Theta^d=3.0$ a.u. The peak value of the strongest field (Fig.2c) is $\approx 0.2$ a.u. (corresponding intensity is $\approx 10^{14} Watt/cm^2$) which can be readily attained in the laboratory. However, the pulse shape presents a chalange to the present-day laser shaping technology. A detailed characterization of the optimal field can however, be made by Fourier transforming the fields. Fig. 3 shows the average distance $<x>$ as a function of time, the corresponding control fields are shown in Fig.2. Notice that the molecular phase, $\phi_M$ changes the course of the dynamics of excitation as evidenced by the behaviour of the the expectation value of $<x>$ and the corresponding optimal control fields. 

Figure 4 shows the optimal electric field and the corresponding expectation value of x for the superposition $|000>+|200>$. This result can be compared to that of Figs. 2a and 3a for the superposition $|000>+|100>$. Figure 5 on the otherhand, shows $<x>(t)$ and the corresponding optimal field for different values of $|c_2|^2$ for the superposition $|000>+|100>$. 

\section{Conclusion}

This paper presents the optimal design of the electric field by using the QFD formulation. Molecular coherence has been introduced in such design of the electric field, by creating a nonstationary superposition state comprising of two vibrational eigenfunctions of HBr molecule in its ground electronic configuration, prior to its submission to the electric field to be designed. The molecular coherence is created by allowing the molecule to interact with a superposed laser field. We showed that the molecular phase $\phi_M$ can be experimentally varied and used as a parameter to modify the optimal electric field so as to manipulate certain dynamical change in the molecule. Applications to the multiphoton excitation of HBr molecule, described by the expectation value of x, show that control range is extensive. The results show that pulses with different structures result in different excitation processes since they follow different interfering routes, and the interference routes are controlled by the molecular phase. One can indeed optimize other quantity with some constraints in it so as to follow only one out of several interference routes.

The extent to which molecular coherence (and hence, the quantum interference) enters into the optimal results of the elctric field is central to the understanding of the control of the excitation. At present we found a variety of results. In several cases such interference was indispensable in producing the optimal results, often necessitaing the involvement of large numbers of interferring routes to the excited states. Although the multiphoton excitation phenomena are not so suitably described by the operator x, this is the first case study where we combined the coherent control method with the optimal control method. Other multiphoton excitation processes, with suitable operators, are currently under investigation.

Note that one could use the standard control equations based on the TDSE, however, recent studies \cite{ref20,ref24,ref,25} show that the QFD equations require lesser number of spatial grid points than the TDSE when solved numerically which apparently enhances the efficiency and the numerical saving of the QFD-OCT. This enhancement is attributed mainly to the relatively slow varying nature of the hydrodynamical variables, compared to the wave function, in the spatial coordinate. It may be mentioned that for other target operators the TDSE based control equations may be easier to handel than the QFD based equations however, in the present study the chosen operator is such that QFD based equations are more easy to solve.

\pagebreak

{\bf Figure Captions}

Figure 1 : Time(in picosecond) variation of $|c_2(t)|^2$ (a), $\Re(c_2(t))$ (b), $\Im(c_2(t))$ (c) and $\phi _M(t)$ (d) corresponding to $c_1|000>e^{-\omega_1t}+c_2(t)|100>e^{-\omega_2t}$ superposition state, where $\Re$ and $\Im$ refer to real and imaginary respectively. This superposition is obtained with a short(spectrally) laser pulse whose central frequency corresponds to 432.61 nm, FWHM corresponds to 5.54 ps = 5.31 $cm^{-1}$ $t_0=0$ and the peak intensity is $2.647\times10^6 Watt/cm^2$.

\vspace{0.2in}
Figure 2: Optimal electric pulse, E(t) in atomic unit (a.u.) plotted against time in a.u. Label (a), (b), (c) and (d) correspond to the molecular phase , $\phi _M$ = $\pi/5$, $-\pi/5$, $\pi/3$ and $\pi$ radian respectively. The superposition state is $c_1|000>e^{-iE_1t/\hbar}+c_2(t)|100>e^{-iE_2t/\hbar}$ where $|c_1|^2=0.8$ and $|c_2|^2=0.2$ and $\Theta^d=3.2$ a.u.

\vspace{0.2in}
Figure 3: Average distant, $<x>$ in atomic unit (a.u.) plotted against time in a.u. Label (a), (b), (c) and (d) correspond to the molecular phase , $\phi _M$ = $\pi/5$, -$\pi/5$, $\pi/3$ and $\pi$ radian respectively. The superposition state is $c_1|000>e^{-iE_1t/\hbar}+c_2(t)|100>e^{-iE_2t/\hbar}$ where $|c_1|^2=0.8$ and $|c_2|^2=0.2$ and $\Theta^d=3.2$ a.u. The corresponding electric fields are shown in Fig.2.

\vspace{0.2in}
Figure 4: Optimal electric field (label (a)) and average distant (label (b)) in atomic unit (a.u.) plotted against time in a.u. for $\phi _M=\pi/5$. The superposition state is $c_1|000>e^{-iE_1t/\hbar}+c_2(t)|200>e^{-iE_2t/\hbar}$ where $|c_1|^2=0.8$ and $|c_2|^2=0.2$ and $\Theta^d=3.2$ a.u.

\vspace{0.2in}
Figure 5: Optimal electric field (label (a) and (c)) and the corresponding average distant (label (b) and (d)) plotted against time for $\phi _M=\pi/5$ radian and $\Theta^d=3.2$ a.u. The superposition state is $c_1|000>e^{-iE_1t/\hbar}+c_2(t)|100>e^{-iE_2t/\hbar}$ where $|c_2|^2$=0.3 (label (a) and (b)) and 0.1 (label (c) and (d)).

\begin{figure}[!h]
\epsfxsize=6.0in
\hspace*{0.5cm}\epsffile{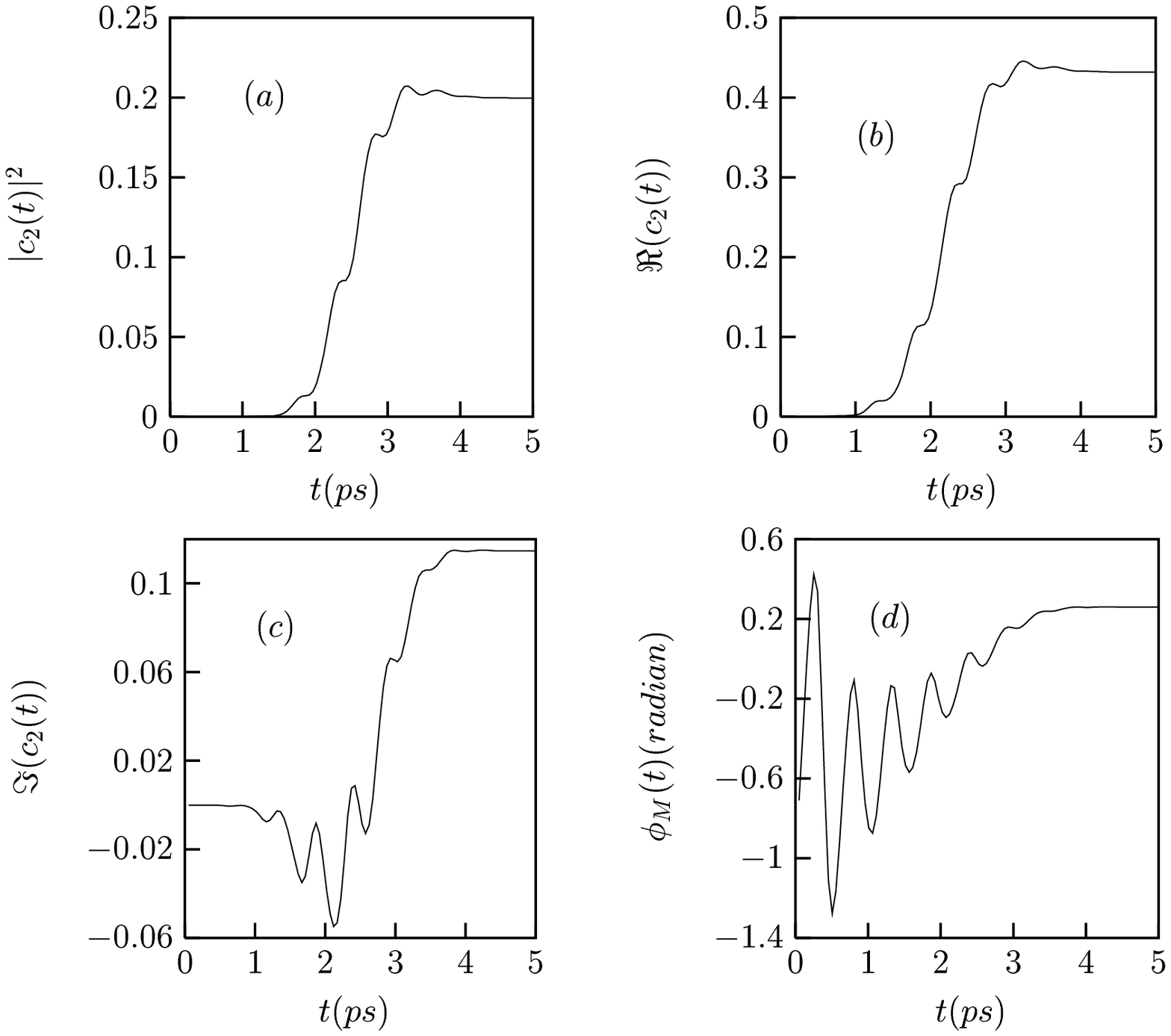}
\vspace*{-0.5cm}\caption{}
\label{c2}
\end{figure}

\begin{figure}[!h]
\epsfxsize=6.0in
\hspace*{0.5cm}\epsffile{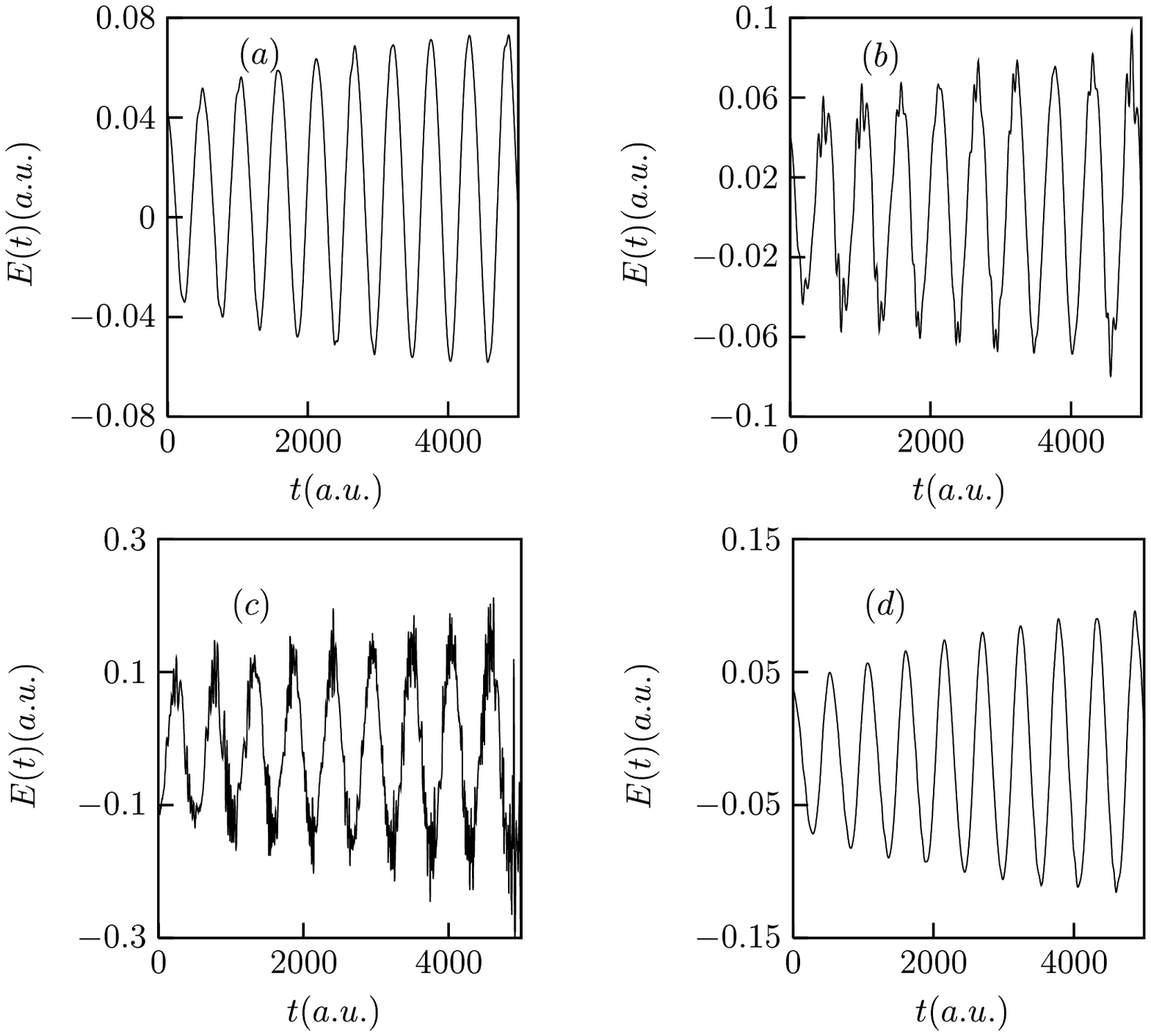}
\vspace*{-0.5cm}\caption{}
\label{Et}
\end{figure}

\begin{figure}[!h]
\epsfxsize=6.0in
\hspace*{0.5cm}\epsffile{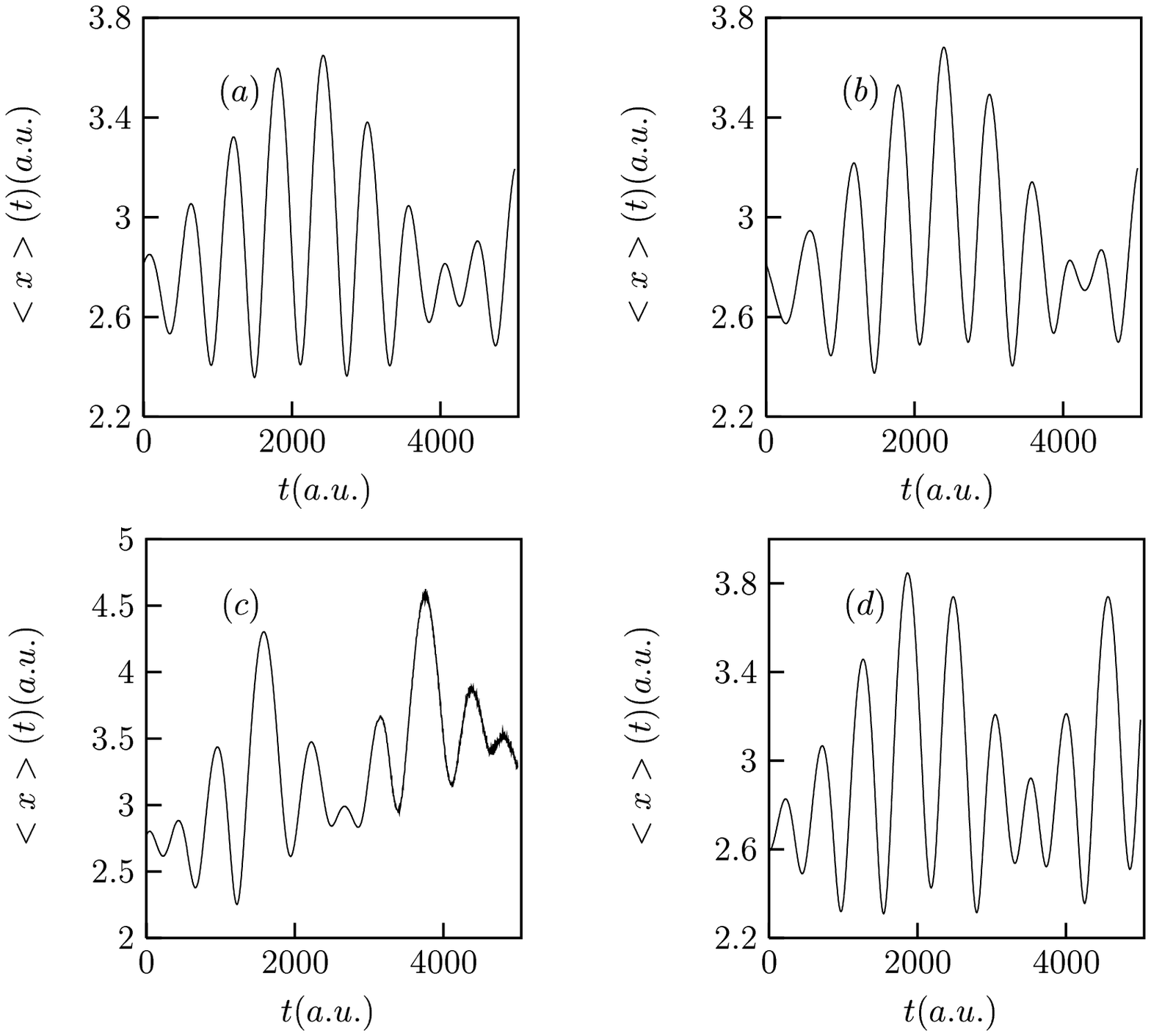}
\vspace*{-0.5cm}\caption{}
\label{xt}
\end{figure}

\begin{figure}[!h]
\epsfxsize=6.0in
\hspace*{0.5cm}\epsffile{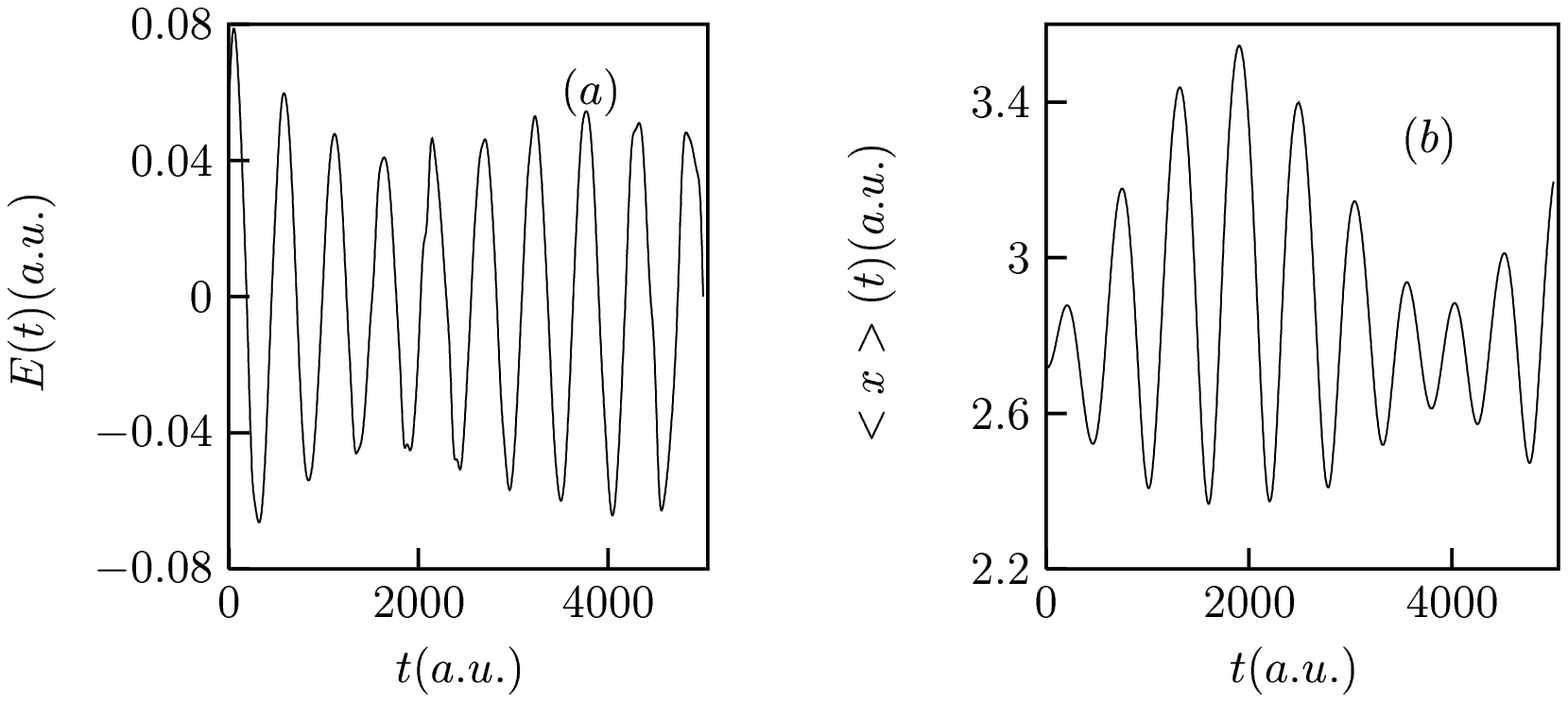}
\vspace*{-0.5cm}\caption{}
\label{Etxt1}
\end{figure}

\begin{figure}[!h]
\epsfxsize=6.0in
\hspace*{0.5cm}\epsffile{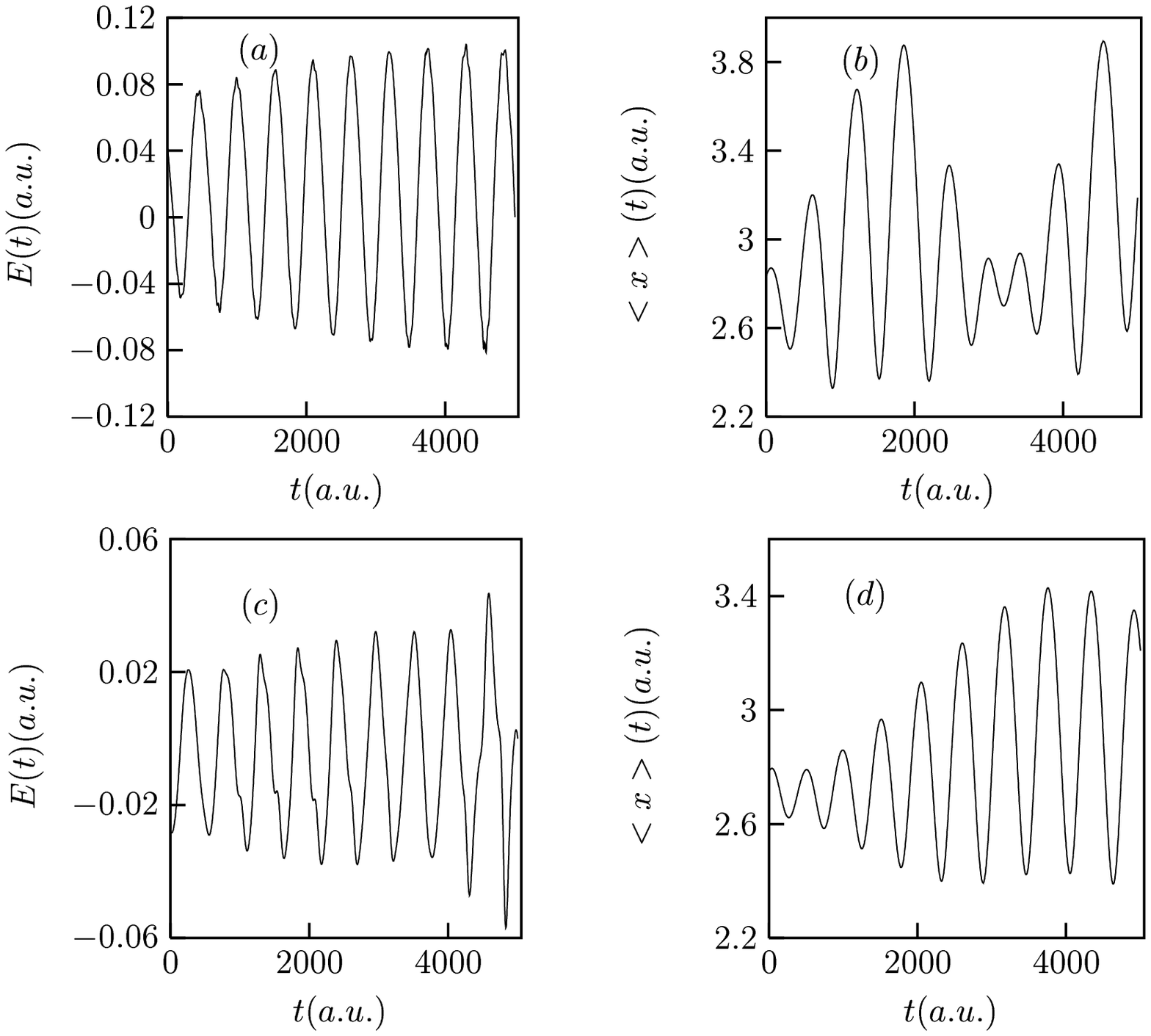}
\vspace*{-0.5cm}\caption{}
\label{Etxt1}
\end{figure}


\newpage
\begin{references}

\bibitem{ref1}S. A. Rice, Science, {\bf 258}, 412, 1992.

\bibitem{ref2}D. J. Tannor and S. A. Rice, J. Chem. Phys., {\bf 83}, 5013, 1985.
\bibitem{ref3}A. P. Peirce, M. A. Dahleh and H. Rabitz, Phys. Rev. {\bf A 37}, 4950, 1988.

\bibitem{ref4}W. S. Warren, H. Rabitz and M. Dahleh, Science, {\bf 259}, 1581, 1993.

\bibitem{ref5}P. Brumer and M. Shapiro, Faraday Discuss. Chem. Soc., {\bf 82}, 177, 1986.

\bibitem{ref6}P. Brumer and M. Shapiro, Annu. Rev. Phys. Chem., {\bf 43}, 257, 1992.

\bibitem{ref7}B. Kohler, J. Krause, F. Raksi, K. R. Wilson, R. M. Whitnell, V. V. Yakovlev and Y. J. Yan, Acct. Chem. Res., {\bf 28}, 133, 1995.

\bibitem{ref8}T. Baumert and G. Gerber, Isr. J. Chem., {\bf 34}, 103, 1994.

\bibitem{ref9}H. Rabitz and S. Shi, Adv. Mol. Vib. Collision Dyn. {\bf 1A}, 187, 1991.

\bibitem{ref10}R. S. Judson and H. Rabitz, Phys. Rev. Lett., {\bf 68}, 1500, 1992.

\bibitem{ref11}A. Assion, T. Baumert, M. Bergt, T. Brixner, B. Kiefer, V. Strehle and G. Gerber, Science, {\bf 282}, 919, 1998.

\bibitem{ref12}C. J. Bardeen, V. V. Yakovlev, K. R. Wilson, S. D. Carpenter, P. M. Weber and W. S. Warren, Chem. Phys. Lett., {\bf 280}, 151, 1997.

\bibitem{ref13}A. Assion, T. Baumert, V. Seyfried and G. Gerber, in Ultrafast Phenomena, edited by J. Fujimoto, W. Zinth, P. E. Barbara and W. H. Knox, springer, Berlin, 1996.

\bibitem{ref14}A. Shnitman, I. Sofer, I. Golub, A. Yogev, M. Shapiro, Z. Chen and P. Brumer, Phys. Rev. Lett., {\bf 76}, 2886, 1996.

\bibitem{ref15}D. J. Tannor, R. Kosloff and S. A. Rice, J. Chem. Phys., {\bf 85}, 5805, 1986.

\bibitem{ref16}T. Baumert, M. Grosser, R. Thalweiser and G. Gerber, Phys. Rev. Lett., {\bf 67}, 3753, 1991.

\bibitem{ref17}E. D. Potter, J. L. Herek, S. Pedersen, Q. Liu, A. H. Zewail, Nature, {\bf 355}, 66, 1992.

\bibitem{ref18}S. Shi, A. Woody, H. Rabitz, J. Chem. Phys., {\bf 88}, 6870, 1988.

\bibitem{ref19}B. K. Dey, H. Rabitz and A. Askar, Phys. Rev. {\bf A61}, 043412, 2000.

\bibitem{ref20}B. K. Dey, J. Phys., {\bf A33}, 4643, 2000.

\bibitem{ref21}S. H. Lin, Y. Fujimura, H. J. Neusser and E. W. Schlag, Multiphoton Spectroscopy of Molecules, Academic Press, Inc., Ch.4, p 89, 1984.

\bibitem{ref22} E. F. Van Dishoeck, M. C. Van Hemert and A. Dalgarno, J. Chem Phys., {\bf 77} 3693, 1982.

\bibitem{ref23} E. McCullough, M. Shapiro and P. Brumer, Phys. Rev. {\bf A61}, 04180, 2000.

\bibitem{ref24}B. K. Dey, A. Askar and H. Rabitz, J. Chem. Phys., {\bf 109}, 8770, 1998.

\bibitem{ref25}B. K. Dey, A. Askar and H. Rabitz, Chem. Phys. Lett., {\bf 297}, 247, 1998.

\bibitem{ref26}G. G. Balint-Kurti, C. L. Ward and C. C. Martson, Comput. Phys. Commun, {\bf 67}, 285, 1991.

\bibitem{ref27}K. P. Huber and G. Herzberg, Molecular Spectra and Molecular Structure IV. Constants of Diatomic Molecules, New York, Van Nostrand Reinhold, 1979.
\end{references}
\end{document}